# Reduction of Fermi velocity in folded graphene observed by resonance Raman spectroscopy

Zhenhua Ni[#], Yingying Wang[#], Ting Yu, Yumeng You, and Zexiang Shen[*]

*Division of Physics and Applied Physics, School of Physical and Mathematical Sciences, Nanyang Technological University, Singapore 637371, Singapore*

**Abstract**

The 1+1 layer folded graphene sheets that deviate from AB stacking are successfully fabricated and their electronic structures are investigated by Raman spectroscopy. Significant blue shift of the 2D band of folded graphene compared to that of single layer graphene (SLG) is observed. This is attributed to SLG-like electronic structure of folded graphene but with slowing down of Fermi velocity (as much as ~5.6%). Different amount of blue shift of 2D band is observed for different folded graphenes, which may correspond to the different twist angle and/or separation between the two layers, resulting in different Fermi velocity of folded graphenes. Electronic structure of 1+1 folded graphene samples with different stacking order (twist and separation between the two layers, and in-plane shift of the two layers) can be investigated by Raman spectroscopy.

**Introduction**

Graphene, since firstly fabricated by micromechanical cleavage of graphite in 2004,[1] has revealed a lot of unusual properties, such as the ballistic transport,[2] anomalous quantum hall effect,[3] which are closely related to its unique electronic band structure. For single layer graphene (SLG), it is known that the low energy dispersion is linear,[2] which make charge carriers in SLG behave like massless Dirac fermions. The properties of bi-layer graphene (BLG) have been interpreted under the assumption that the stacking of the two graphene layers has the form of A, B (Bernal) stacking. As a result, the $\pi$-electrons dispersion in the valence and conduction bands splits into two parabolic branches near the K point.[4] The electronic band structure of multi-layer graphene is closely related to the stacking order of graphene layers.

Interestingly, recent studies such as magnetotransport,[5] far infrared magneto-transmission[6] investigations on multi-layer epitaxial graphene (EG) grown on SiC substrate still revealed the two dimensional Dirac-like (SLG-like) character of electronic states. It is verified by STM (scanning tunneling microscopy)[7] and LEED (low energy electron diffraction)[8,9] that there is a high degree of rotational disorder in the multi layer EG. Such stacking disorder makes the electronic structure of SLG preserved even in multilayer EG. Furthermore, theoretical calculations of electronic structure of BLG with a twist of second layer were carried out. The results show that the low energy dispersion of twisted two-layer graphene is linear, as in SLG, but the Fermi velocity is significantly smaller.[10] Further investigation on the electronic structure of BLG that deviates from the AB stacking is necessary to obtain fundamental understanding of the relation between stacking order and electronic properties of bi- or multi- layer graphene.

In this paper, we have successfully fabricated the 1+1 folded graphene with different stacking order and studied their electronic properties using Raman spectroscopy. There are two characteristic bands in the Raman spectrum of graphene, the in-plane vibrational G band (at ~1580 cm$^{-1}$) and the two-phonon double resonant 2D band (~2670 cm$^{-1}$).[11-14] Taking advantage of the double resonance effect, the relation between electronic structure and Raman 2D band is established. This 2D Raman band has been widely used in the studies of graphene, such as identifying the number of graphene layers [11-14] and probing electronic structure.[15] In our results, obvious blue shift of the 2D band of folded graphene is observed which is interpreted in term of change of electronic structures. It is believed that the electronic structure of folded graphene is similar to SLG but with smaller Fermi velocity which agrees very well with the theoretical simulation by Lopes dos Santos *et al*.[10] Finally, the Fermi velocity of folded graphene is estimated according to the blue shift of 2D band from the double resonance process.

**Experimental methods**

The graphene samples are prepared by micromechanical cleavage and transferred to Si wafer with ~300 nm SiO$_2$ capping layer.[1] Our 1+1 layer samples are prepared by simply gently flushing de-ionized water across the surface of the substrate containing the target graphene sheet. The 1+1 layer folded graphenes can be observed after this process. We have made a total of six folded samples using this method and the size of the folded area is between 5 to 10 um$^2$.. Raman imaging /spectroscopy are carried out with a WITEC CRM200 Raman system with 532 nm (2.33 eV) excitation and laser power at sample below 0.1 mW to avoid laser induced heating. A 100× objective lens with a NA=0.95 is used in the Raman experiments. To obtain the Raman images, a piezo stage is used to move the sample with step size of 200 nm and Raman spectrum is recorded at every point. The stage movement and data acquisition are controlled using ScanCtrl Spectroscopy Plus software from WITec GmbH, Germany. Data analysis is done using WITec Project software.

**Results and discussion**

Fig. 1(a) and (b) show the optical images of SLG before and after folding. SLG is identified by Raman spectrum according to the characteristic of 2D band (very sharp with bandwidth ~30 cm$^{-1}$ and symmetric). [11-14] The thickness is further verified by optical contrast spectrum. [16] The black dashed rectangle indicates the location where folding happens. The size of graphene is about 10x10 um$^2$ and the folded area is about several um$^2$. As the optical images are not very clear, Fig. 1(c) and (d) show the schematic drawing of the sample before and after folding. Fig. 1(e) and (f) respectively show Raman images of 2D and G band position (frequency) after folding. Brighter color represents higher frequency of Raman bands. In Fig. 1(e), different contrast can be observed, which is attributed to different 2D band frequency. It is obvious that the 2D band frequency of folded graphene is much higher than that of

SLG. On the other hand, in Fig. 1(f), even contrast can be seen which means almost no change of G band frequency before and after folding. Fig. 1(g) shows the Raman image obtained from the 2D peak area (integrated intensity) after folding. Brighter color represents higher 2D band intensity. The intensity of folded graphene is much higher than that of SLG. It is observed in the previous work [14] that the area of the 2D band is almost identical for 1 to 4 layers graphene. Hence, the much stronger 2D band of folded graphene reveals its structure and properties are different from that of BLG. Fig. 1(h) gives the Raman image extracted from the G band area after folding. The G band intensity of folded graphene is nearly double of SLG since it contains two layers of carbon atoms. Note that the 2D band intensity increases much more than that of G band after folding. Statistical analysis of the Raman images is summarized in Table 1. The analyzed areas **X** (folded graphene) and **Y** (SLG) are marked by black squares in Fig. 1(e). It can be seen that ~ 12cm$^{-1}$ blue shift of 2D band is observed after folding. The 2D peak area of folded graphene is almost 3 times as high as that of SLG. More surprisingly, its width is smaller than that of SLG.

Fig. 2 gives typical Raman spectra of the 1+1 folded graphene (taking from area **X** in Fig. 1(e)), SLG (taking from area **Y** in Fig. 1(e)) and BLG for comparison. The spectra are normalized to have similar G band intensity. From this figure, it can be obviously seen that the Raman spectrum of 1+1 folded graphene is different from that of BLG.. The 2D band of BLG is much broader and can be fitted as 4 peaks, which originates from splitting of valence and conduction bands.[14] However, for folded graphene, only a single sharp peak exists which is similar to that of SLG. Thus, the electronic structure of folded graphene should be similar to that of SLG, i.e. there is no splitting of energy bands. Although the Raman features of folded graphene is quite similar as that of SLG, there are differences needed to be noticed: A strong blue shift (~12 cm$^{-1}$) of the 2D band of folded graphene can be clearly seen, as indicated in the Raman images (Fig. 1(e)). This blue shift is associated with the SLG-like band structure of 1+1 folded graphene but with smaller Fermi velocity, which will be discussed in detail latter. The folded graphene has higher 2D to G band intensity ratio than that of SLG, partially due to the different resonance conditions of folded graphene and SLG.

Fig. 3(a) and (b) show the optical images of another 1+1 folded graphene sheet before and after folding. The white dashed rectangle indicates the area where folding occurs. The sample contains both SLG (size of ~ 20 x 5 um$^2$) and BLG (size of ~20 x 5 um$^2$), as indicated in the figures. The size of the 1+1 folded area is about 5 um$^2$. Figs. 3(c) and (d) give the schematic images of the graphene sample before and after folding, while Figs. 3 (e) and (f) give the 2D band intensity Raman image of sample before and after folding. Before folding, the 2D band area is almost the same for SLG and BLG (evenly distributed contrast in Fig.3(e)).[14] However, the 2D band areas of 1+1 folded graphene as well as 1+2 folded graphene are much higher than that of SLG/BLG.. This indicates the different electronic structures of folded graphene (1+1 and 1+2) from that of normal BLG and three layer graphene. Figs. 3(g) and (h) are the 2D frequency images of graphene before and after folding, brighter color represents higher frequency. It can be seen that before folding, the 2D band frequency is roughly constant for SLG. However, after folding, the 1+1 folded graphene has higher 2D band frequency compared to that of SLG. Statistical analysis of Raman images is summarized in Table 2. The included areas **X** (1+1 layer folded graphene) and **Y** (SLG) are marked by black squares in Fig. 3(f). From Table 2, it can be seen that the blue shift of the 2D band of folded graphene is ~ 4 cm$^{-1}$, which is smaller than the previous folded graphene. The 2D band is broader after folding, which is also different from the previous sample. The exact cause for the change of 2D band width of folded graphene is still not clear. It may also be related to the different resonance conditions of folded graphene and SLG. We have studied a total of six 1+1 layer folded graphene samples. Blue shift of the 2D band compared to SLG is observed for every sample, but the amount of blue shift differs from 4 to 14 cm$^{-1}$. Four of the folded samples show narrowing of the 2D band width, while two of them show broadening.

Lopes dos Santos *et al.* [10] calculated the electronic structure of twisted BLG and their results show that it has SLG-like linear dispersion with slowing down of Fermi velocity. The stacking order of our folded 1+1 sample is surely different from the normal BLG.. It has different rotation angles of second layer, which can be viewed as twisted BLG with unknown twisted angle. Therefore, the energy band structure of our folded graphene is estimated under the model of twisted BLG.. Fig. 4 schematically shows the electronic structure of 1+1 folded graphene as well as SLG.. The slowing down of Fermi velocity ($v_F$') of folded graphene corresponds to the smaller slope of energy dispersion near Dirac point. The second-order double resonance Stokes process for folded graphene and SLG are schematically shown in Fig. 4. In double resonance process, the excitation photon with energy ~2.33 eV firstly creates an e-h pair with similar energy at wave vector *k*. Following, electron-phonon scattering happens with an exchanged momentum of *q/q'* for SLG and folded graphene. Here, *q* is the phonon momentum for SLG and *q'* is the phonon momentum for folded graphene. After that, electron-phonon scattering with an exchanged momentum with reverse direction *-q/q'* happens, followed by the e-h recombination. The wave vector *q/q'* decides the 2D band frequency. It can be seen from Fig. 4 that for folded graphene, phonon with higher wave vector (*q'*) is needed to inelastically scatter the electron compared to that of SLG (*q*). Because of the almost linear dispersion of optical phonon branch around K point ($\frac{d\omega}{dq}$) [13] which contributes to the observed 2D band frequency, phonon with higher frequency (ω') is obtained in double resonance process for folded graphene. This is the reason of blue shift of the 2D band of folded graphene. By using this model, the sharpness and blue shift of 2D band of multilayer EG on carbon terminated SiC can be also easily understood, [17] which are caused by the SLG-like linear dispersion of energy band of multilayer EG but with ~10% slowing down of Fermi velocity [6,18] than that of SLG.

Moreover, the Fermi velocity of folded graphene can be estimated. In the double resonance process, [19] it is known that

$$\hbar v_F q \approx E_L - \hbar(\omega/2) \quad (1)$$

Here, $E_L$ is the incident photon energy, $v_F$ is the Fermi velocity, ω is the frequency of two phonon 2D band, $\hbar(\omega/2)$ is the phonon energy.

Therefore,

$$qdv_F + v_F dq = -1/2 \cdot d\omega \tag{2}$$

Hence,

$$q\frac{dv_F}{v_F} + dq = -\frac{d\omega}{2v_F} \tag{3}$$

$$\frac{dv_F}{v_F} = -(1/2 + \frac{v_F}{d\omega/dq})\frac{1}{v_F q}d\omega \tag{4}$$

Where $\frac{d\omega}{dq} = 645 cm^{-1} \overset{\circ}{A} = 0.08 eV \overset{\circ}{A}$ is the phonon dispersion around K point.[11] We assume it is same as SLG as the G band frequency does not change after folding.

Because $\frac{v_F}{d\omega/dq} \gg 1/2$, the equation can be written as:

$$\frac{dv_F}{v_F} = -\frac{1}{q\frac{d\omega}{dq}}d\omega = -\frac{\hbar v_F}{[E_L - \hbar(\omega/2)] \cdot \frac{d\omega}{dq}}d\omega \tag{5}$$

Here, $\hbar v_F = 6.5 eV \overset{\circ}{A}$,[19] $E_L$=2.33eV, $\hbar(\omega/2) = 0.166 eV$ by taking ω≈2670 cm⁻¹.

Therefore,

$$\frac{dv_F}{v_F} = -\frac{37.5 \cdot d\omega}{eV} or -\frac{0.00467 \cdot d\omega}{cm^{-1}} \tag{6}$$

where $\frac{dv_F}{v_F}$ is the Fermi velocity change in percentage and $d\omega$ is the frequency change of 2D band after fold in eV or cm⁻¹.

For the first 1+1 folded graphene sample, the blue shift of 2D band is ~12 cm⁻¹, thus, the slowing down of Fermi velocity is about 5.6%. For the second sample, the smaller blue shift (~4 cm⁻¹) is corresponding to the smaller slowing down of Fermi velocity (~2%). This difference may be due to the different twist angle between the first and second graphene layer for different samples. Different twist angle between the two layers will result in different amount of slowing down of Fermi velocity,[10] hence the different blueshift of 2D band. Another reason is the separation of the two layers, which will also affect the electronic properties of folded graphene. For the second sample, one edge of SLG is folded on top of BLG as can be seen in Fig. 3(f). This makes the separation between the two layers of 1+1 graphene larger than the first folded sample. The different amount of blue shift for different samples indicates that the different stacking order (twist angle of second layer, separation between two layers, the in-plane shift of the two layers) results in different electronic structure of folded graphene, which agrees with the calculation by Lopes dos Santos *et al*.[10] Raman study of 1+1 folded graphene is an easy way to probe the electronics structure of BLG deviated from A, B stacking. Electronic structure of folded graphene with variable separation and twist angle will be further investigated using Raman spectroscopy.

**Conclusion**

Raman spectroscopy is used to investigate the electronic structure of 1+1 folded graphene and compared with that of SLG.. The 1+1 folded graphene deviated from AB stacking and it is believed that the electronic structure is SLG-like but with slowing down of Fermi velocity. For different stacking order (twist angle and separation between the two layers, as well as the in-plane shift of the two layers), different amount of Fermi velocity slow down (as much as ~5.6%) is obtained from the blue shift of 2D band, which agrees well with the previous theoretical calculation.[10] Our results clearly indicate that resonance Raman spectroscopy can be used to monitor the electronic band structure, i.e. Fermi velocity, of multilayer graphene with different stacking order.

**Figure captions**

(Color online) Fig. 1(a) and (b) show the optical image of SLG before and after fold. (c) and (d) give the schematic image of SLG before and after fold. (e) and (f) respective shows Raman image obtaining from the 2D and G band positions. (g) and (h) individually show Raman image by extracting the area of 2D and G band.

(Color online) Fig. 2 the Raman spectrum of BLG, SLG and 1+1 layer folded graphene. The spectra are normalized to have similar G band intensity.

(Color online) Fig. 3(a) and (b) show the optical image graphene layers before and after fold. (c) and (d) give the schematic images of SLG before and after fold. (e) and (f) give the Raman images of graphene by extracting the area of the 2D band before and after fold. (g) and (h) show Raman images by extracting the position of the 2D band before and after fold.

(Color online) Fig.4 schematically shows the electronic structure of 1+1 layer folded graphene (black dash lines) as well as SLG (purple solid lines). The arrows indicate the double resonance process in folded graphene (black dash lines) and SLG (red solid lines).

TABLE 1 Statistical analysis results of Raman image of 1+1 layer folded graphene (area X) as well as SLG (area Y) shown in Fig.1.

|  | Position (cm$^{-1}$) | | Area (arb. units) | | Width (cm$^{-1}$) | |
| --- | --- | --- | --- | --- | --- | --- |
|  | X | Y | X | Y | X | Y |
| 2D band | 2686.6±0.5 | 2674.4±1.0 | 202.8±21.5 | 87.8±18.3 | 22.5±1.1 | 24.9±2.5 |
| G band | 1581.4±1.0 | 1582.3±1.5 | 34.4±7.4 | 20.0±5.7 | 12.8±2.9 | 11.7±4.4 |

TABLE 2 Statistical analysis results of Raman image of 1+1 layer folded graphene (area X) as well as SLG (area Y) shown in Fig.3.

|  | Position (cm$^{-1}$) | | Area (arb. units) | | Width (cm$^{-1}$) | |
| --- | --- | --- | --- | --- | --- | --- |
|  | X | Y | X | Y | X | Y |
| 2D band | 2674.6±0.4 | 2670.4±0.8 | 569.6±20.1 | 295.0±20.6 | 31.8±1.1 | 27.7±1.1 |
| G band | 1581.8±1.3 | 1580.6±1.0 | 98.2±10.4 | 51.9±8.3 | 17.2±1.6 | 16.7±2.9 |

Fig. 1

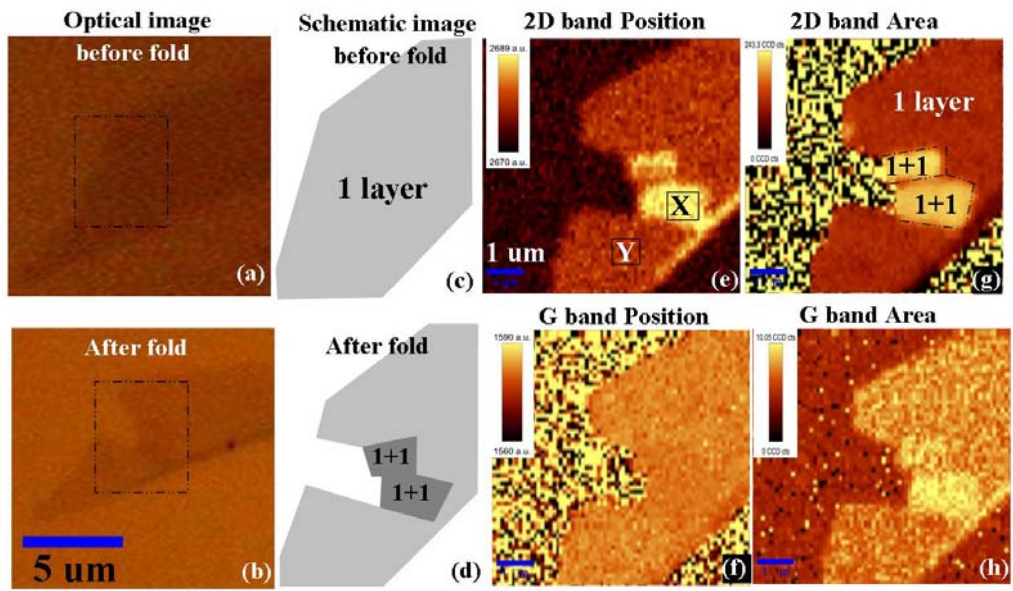

Fig. 2

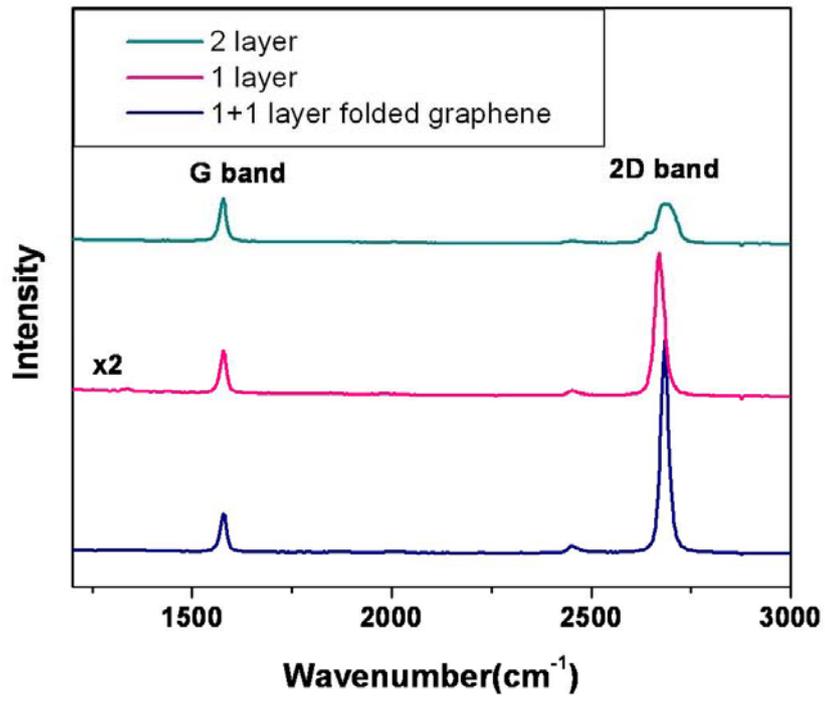

Fig. 3

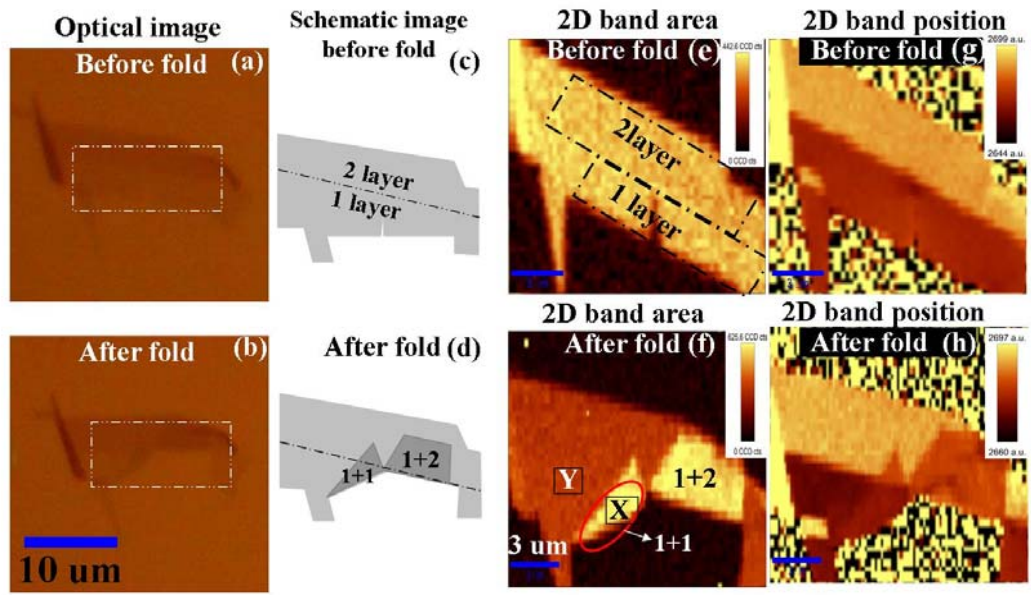

Fig. 4

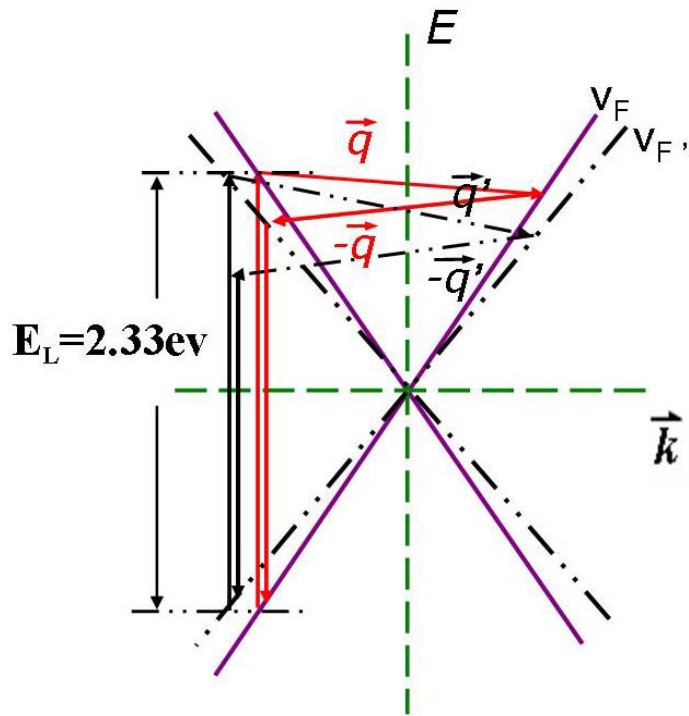